\documentclass[fullpaper,final]{nldl}

\paperID{59}

\vol{233}

\usepackage{mathtools}

\usepackage{graphicx}

\usepackage{booktabs}

\usepackage{enumitem}

\usepackage{algorithm}
\usepackage{algorithmic}

\usepackage{multirow}

\usepackage{listings}
\usepackage{hyperref}
\usepackage{url}
\hypersetup{
  pdfusetitle,
  colorlinks,
  linkcolor = BrickRed,
  citecolor = NavyBlue,
  urlcolor  = Magenta!80!black,
}

\title{Local Gamma Augmentation for Ischemic Stroke Lesion Segmentation on MRI}
\author[1,3]{Jon Middleton$^\dag$\thanks{Corresponding Author: jami@di.ku.dk}}
\author[3]{Marko Bauer\thanks{Equal contribution.}}
\author[1,3]{Jacob Johansen}
\author[3]{Mathias Perslev}
\author[1,3,4]{Kaining Sheng}
\author[3,4]{Silvia Ingala}
\author[1,2,3]{Mads Nielsen}
\author[3,4]{Akshay Pai}
\affil[1]{Department of Computer Science, University of Copenhagen}
\affil[2]{Pioneer Centre for Artificial Intelligence, University of Copenhagen}
\affil[3]{Cerebriu A/S}
\affil[4]{Department of Diagnostic Radiology, Rigshospitalet, Copenhagen, Denmark}

\begin{document}
\maketitle

\begin{abstract}  
The identification and localisation of pathological tissues in medical images continues to command much attention among deep learning practitioners. When trained on abundant datasets, deep neural networks can match or exceed human performance. However, the scarcity of annotated data complicates the training of these models. Data augmentation techniques can compensate for a lack of training samples. However, many commonly used augmentation methods can fail to provide meaningful samples during model fitting. We present local gamma augmentation, a technique for introducing new instances of  intensities in pathological tissues. We leverage local gamma augmentation to compensate for a bias in intensities corresponding to ischemic stroke lesions in human brain MRIs. On three datasets, we show how local gamma augmentation can improve the image-level sensitivity of a deep neural network tasked with ischemic stroke lesion segmentation on magnetic resonance images.
\end{abstract}

\section{Introduction}
The potential of deep neural networks to identify and localise pathological tissues in medical images continues to motivate research into novel  training methods. In theory, deep architectures can achieve arbitrarily high levels of performance. In practice, however, these algorithms are trained on insufficient quantities of data. Overfitting arises as a result, and the full  potential of these algorithms remains unrealised.

Traditional approaches to the data scarcity problem use data augmentation to expand the size of the dataset. A typical setup selects a collection of data-agnostic operations which come from  parameterized families of spatial and intensity transformations. Examples include affine transformations, elastic transformations, global gamma transformations, additive and multiplicative noise transformations, blurring transformations, and brightness transformations. An unlimited collection of novel samples can then be obtained by fixing probability distributions on the parameter spaces and sampling augmentation parameters from them.

Though a dataset can be enlarged in this fashion, the approach may not provide relevant samples for image models tasked with classifying or segmenting regions of interest. Moreover, the augmentations themselves are not guaranteed to have an appreciable impact on model training. For example, the literature often claims that elastic deformations improve deep neural network performance \cite{Basaran2022a,Castro2018,Cirillo2021a,Novosad2020}. Nonetheless, these transformations do not appear in all well-known augmentation pipelines (for example, \cite{Isensee}), and some research indicates that these transformations can impair a network's performance \cite{Karimi2023,Perez2018}. The latter examples fit into the claim in \cite{Balestriero2022} that the effectiveness of a data augmentation pipeline depends crucially on the content of a dataset.

Thus, for models tasked with identifying and localising pathological tissues in medical images, practitioners are developing augmentation techniques that target these tissues. Recent work \cite{Basaran2022a,Fernandez-Quilez2023,Liu2022c,Perez2018,Zhang2023b} has focused on local augmentations that expose deep networks to additional examples by isolating and altering regions of interest.

We contribute a simple, targeted method of local data augmentation for a pathology segmentation model and apply this method to the problem of ischemic stroke lesion  segmentation on multi-modal magnetic resonance images. The method exploits both the single-parameter family of gamma transformations and ischemic stroke lesion segmentation maps to mitigate an intensity bias in the training data. We compare local gamma augmentation against a baseline without it, and we show an improvement in image-level sensitivity as a result.

\begin{figure*}[h!tb]
    \centering
    \includegraphics[width=1.0\textwidth]{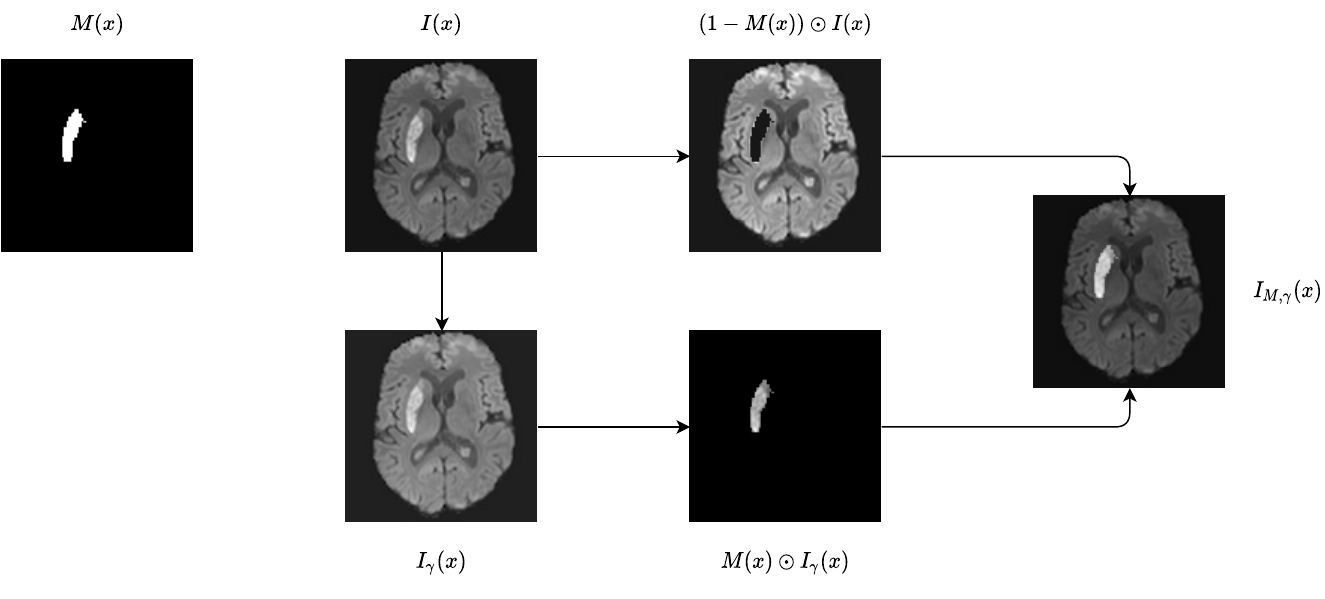}
    \caption{Depiction of local gamma augmentation. An image $I(x)$ and its gamma transformation $I_\gamma(x)$ are multiplied pointwise with a segmentation map and combined to produce an image $I_{M, \gamma}(x)$ featuring an augmentation of a pathology.}
    \label{fig:gamma_pipeline}
\end{figure*}
			
\section{Related Work}
Gamma transformations frequently appear in probabilistic data augmentation pipelines for deep segmentation networks. The training pipeline for nnU-Net \cite{Isensee} uniformly samples parameters for both gamma compression and gamma expansion, with a bias toward gamma expansion parameters. A similar gamma augmentation scheme for a tumour segmentation network appears in \cite{Cirillo2021a}, with equal weighting of gamma compression and gamma expansion. In \cite{Sun2021}, random channel-wise gamma augmentations are applied to RGB images of retinal vessels, with a strong bias toward gamma expansion. In \cite{Liskowski2016}, random gamma compressions augment the saturation and value channels of HSV images of retinal vessels. In all cases, the gamma transformations are applied to the whole image and are uninformed by any features of the dataset. 

Other applications for global gamma augmentations include generative modeling, out-of-distribution benchmarking, and unsupervised domain adaptation. In \cite{Billot2021a}, log-normally distributed gamma transformations assist in the construction of synthetic MRI training data for a segmentation network. In \cite{Nijskens2023}, gamma augmentations help to produce synthetic training data for a contrast-agnostic MRI-to-CT generation network. In \cite{Boone2022}, the intensity distribution shifts induced by gamma transformations are used to assess the robustness of image segmentation networks. Moreover, the authors observe that gamma compression impairs a segmentation network's ability to localise white-matter hyperintensities. In \cite{Diao2022}, a GAN framework is trained to simulate global intensity distribution shifts using instance-specific gamma transformations.

Local image augmentations arise in many training schemes. CarveMix \cite{Zhang2023b} augments MRI training data by inserting pathologies from one image into a second image. Moreover, for brain tumours, the authors use deformable image registration to generate plausible intracranial mass effects. CarveMix is also used in \cite{Basaran2022a} to augment MRIs featuring multiple sclerosis lesions. Mask-based augmentations which corrupt image data, such as random erasing \cite{Zhong2017} and CutOut \cite{DeVries2017}, have been explored in medical image analysis scenarios and show mixed  results. The authors in \cite{Perez2018} show that random erasing, which replaces random image patches with independent and identically distributed point intensities, impairs the performance of a melanoma classification model. In \cite{Liu2022c}, a white matter tract segmentation algorithm is fine-tuned on MRIs in which random subsets of white matter tracts are cut out. In \cite{Fernandez-Quilez2023}, the authors observe that masked image modelling can improve the performance of a prostate lesion classifier.

\section{Methods}

\subsection{Gamma transformations}
Let $I(x)$ be an image with intensity values in the range $[0,1]$. A gamma transformation is defined by a choice of  $\gamma > 0$ via $I(x) \rightarrow I(x)^\gamma.$ Thus a gamma transformation yields a pointwise, non-linear transformation of $I$. 

Gamma transformations generalize to images with arbitrary values by conjugation with min-max normalization:
\begin{equation}\label{eqn:old-gamma}
I(x) \longrightarrow (m_2 - m_1) \left(\frac{I(x) - m_1}{m_2 - m_1}\right)^\gamma + m_1,
\end{equation}
where $m_1 = \min_x I(x)$ and $m_2 = \max_x I(x)$.

A choice of $\gamma$ between 0 and 1 gives a gamma transformation which increases intensities and is called a gamma compression. For $\gamma$ greater than 1, the defined transformation decreases intensities and is called gamma expansion.

\begin{figure*}[h!tb]
    \centering
    \includegraphics[width=1.0\textwidth]{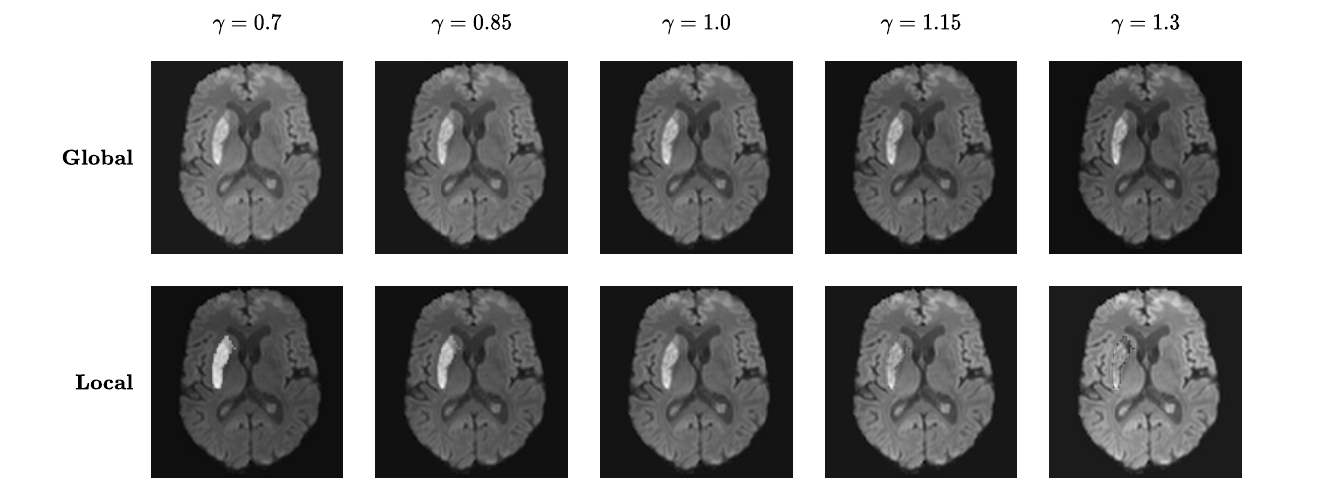}
    \caption{Gamma transformations of a diffusion weighted image from the ISLES-2022 dataset. The center column shows the original image. The top row contains globally gamma compressed images to the left and globally gamma expanded images to the right. The bottom row consists of the same gamma transformations restricted to the location of an acute ischemic stroke.}
    \label{fig:gamma_mris}
\end{figure*}

\subsection{Local gamma augmentations}\label{lga} To adapt gamma transformations to a local region of interest, we alter the min-max normalization in equation \ref{eqn:old-gamma} as follows. Given a binary function $M(x)$, let $S = \{x : M(x)=1\}$ be the set of foreground points of $M$. Set $\tilde m_1 = \min_x S$ and $\tilde m_2 = \max_x S$. Then the map
\begin{equation}
I_\gamma(x) = (\tilde m_2 - \tilde m_1) \left(\frac{I(x) - \tilde m_1}{\tilde m_2 - \tilde m_1}\right)^\gamma + \tilde m_1
\end{equation}
applies a non-linear transformation which is normalized relative to the foreground $S$.

We then complete the local gamma transformation by using $M(x)$ as a mask to mix intensity-augmented pathological tissue in $S$ with normal tissue outside $S$:
\begin{equation}\label{eqn-lga}
I_{M, \gamma}(x) = (1 - M(x)) \odot I(x) + M(x) \odot I_\gamma(x),
\end{equation}
where $\odot$ denotes pointwise multiplication. Local gamma augmentation is depicted in Figure \ref{fig:gamma_pipeline}, with a segmentation map indicating diseased brain tissue.

When used to augment image data, gamma transformations are randomly chosen by sampling $\gamma$ from a probability distribution. Typical choices include the log-normal distribution for a weakly constrained value of $\gamma$ or a Beta distribution supported in a fixed interval. Rather than sampling gamma parameters as in \cite{Isensee}, in which $\gamma$ is drawn from the uniform distribution on the interval $[0.7, 1.5]$, we instead sample from a mixture of uniform distributions:

\begin{equation}
\gamma \sim \frac 1 2\, U(0.7, 1.0) + \frac 1 2\, U(1.0, 1.5).
\end{equation}
Thus we effectively partition this interval into a half-open gamma compression interval $[0.7, 1.0)$ and a closed gamma expansion interval $[1.0, 1.5]$. During each choice of $\gamma$, one of the two intervals is selected with equal probability. 

Figure \ref{fig:gamma_mris} depicts various local and global gamma transformations applied to a diffusion-weighted image from the ISLES-2022 dataset. In this image, one can see a clear increase in contrast between healthy brain tissue and diseased brain tissue for $\gamma < 1$, whereas this contrast is decreased for $\gamma > 1$. This differs from global gamma transformations, in which the local contrast between healthy and pathological tissue remains.

\section{Data}
\subsection{Training Data}
The models were trained on a multi-site in-house brain MRI dataset consisting of 1308 studies, comprising of 420 ischemic strokes, 265 tumors and 234 hemorrhages, as well as 530 normal studies without abnormalities. Each study consisted of three MRI sequences: fluid-attenuated inversion recovery (FLAIR), diffusion-weighted imaging (b0, b1000, and ADC), and either susceptibility-weighted imaging (SWI) or T2* gradient echo (T2*GRE). These sequences were annotated by an in-house medical team and verified by a certified neuroradiologist.

Medical experts classify presentations of ischemic stroke on MRI according to their age \cite{Bernhardt2017}. The various classes include \emph{hyperacute} ischemic stroke, which has the shortest time window from stroke onset. When imaged sufficiently early, hyperacute ischemic stroke lesions appear hyperintense in a DWI sequence and isointense in a FLAIR sequence \cite{Thomalla2011a}. Notably, the MRIs in our training set underrepresent hyperacute ischemic strokes, as shown in Table \ref{tab:pathologies}. This yields a bias in ischemic stroke signal intensities between the DWI and FLAIR sequences.

In addition, we include two publicly available datasets for training. These include the Multimodal Brain Tumor Segmentation (BraTS) \cite{Bakas2017,Bakas2018,Menze2015} Challenge 2019 and Ischemic Stroke Lesion Segmentation (ISLES) Challenge 2022 datasets \cite{Cereda2016,Hakim2021}.

\begin{table}[htbp]
  \centering
  \begin{tabular}{ccc}
    \toprule
    \textbf{Pathology} & \textbf{Type} & \textbf{Count} \\
    \hline
    \multirow{3}{*}{Ischemic Stroke} &   Acute/Subacute   & 342   \\
     &   Hyperacute   & 15    \\
     &  Hemorrhagic   & 63    \\
     \hline
     \multirow{2}{*}{Hemorrhage} & Intra-Axial & 137   \\
      & Extra-Axial  & 97    \\
      \hline
     \multirow{2}{*}{Tumor} &    Intra-Axial   & 95    \\
      &   Extra-Axial    & 170   \\
      \hline
     \multirow{2}{*}{Infection} & Intra-Axial  & 27    \\
      & Extra-Axial  & 2     \\
      \hline
  \end{tabular}
  \caption{Representation of tissue pathologies in training data. Of note is the relative under-representation of hyperacute infarcts.}
  \label{tab:pathologies}
\end{table}

All sequences were acquired in different resolutions and acquisition planes. Accordingly, as part of the pre-processing step, DWI sequences were resampled to isotropic resolution, while FLAIR and SWI/T2* GRE sequences were resampled and co-registered to DWI space.

\begin{table*}[h!tb]
    \centering
    \begin{tabular}{c|cc|cc|cc}
    \toprule
        \multicolumn{1}{c}{} & \multicolumn{2}{c}{\textbf{MVS}} & \multicolumn{2}{c}{\textbf{RH}} & \multicolumn{2}{c}{\textbf{WUS}} \\
        \hline
         & Sensitivity & Specificity & Sensitivity & Specificity & Sensitivity & Specificity\\
        \hline
        U-Net & 0.781 & 0.928 & 0.757 & 0.899 & 0.566 & N/A\\
        U-Net+LG & 0.844 & 0.910 & 0.814 & 0.895 & 0.736 & N/A\\
    \hline
    \end{tabular}
    \caption{Image-level sensitivity and specificity for ischemic stroke detection across three MRI datasets for a baseline U-Net versus a U-Net trained with local gamma augmentation. Specificity for the WUS dataset is non-applicable since all samples in the dataset contain ischemic strokes.}
    \label{tab:image-metrics}
\end{table*}

\subsection{Test Data}
Our methods were evaluated on three independent private datasets, yielding a train-test split that generally arises in a real-world scenario, with a mixture of public and private data for the training set and private data entirely sourced from external sites for testing. The test data consist of DWI (b0 and b1000), FLAIR, and ADC sequences for each patient. At training time, the test data consisted of image-level labels only, rather than the pointwise annotations that are customarily supplied to a segmentation model. The three private datasets are elaborated below:

\textbf{MVS.} A multi-site dataset consisting of 387 studies obtained with Siemens scanners, comprising of 139 ischemic strokes, 93 tumors, 76 hemorrhages, as well as 98 studies with no abnormalities, which were annotated on image level by two neuroradiologists.

\textbf{RH.} A consecutively sampled dataset of 262 acquired brain MRI scans in a routine setting from Rigshospitalet (RH) in Copenhagen, Denmark. The dataset has further been consecutively enriched with positive findings of ischemic strokes (hyperacute, acute, and subacute), brain hemorrhage (intra-axial and extra-axial) and brain tumors (intra-axial and extra-axial) from three other hospital sites in the Capital Region of Denmark. This resulted in a multi-site cohort consisting of 487 scans including 96 ischemic strokes, 80  tumors, 76 hemorrhages, as well as 162 completely normal scans. Furthermore, 73 scans have significant findings that are not attributed to ischemic stroke, hemorrhage, or tumors, e.g., inflammatory disease, pineal cysts, aneurysms, and cavernous malformations. The pathology type of the scans in the dataset was classified by a medical doctor with 3 years’ experience in brain imaging using original radiological report impressions as reference. 

\textbf{WUS.} A collection of brain MRI scans featuring 51 patients who awoke with symptoms of ischemic stroke which were not present before falling asleep. This \emph{wake-up stroke} dataset contains two distinct consecutively sampled cohorts, assembled using distinct stroke protocols,  The first cohort contains 27 adult patients, and the second cohort consists of 24 adult patients. Both cohorts feature visible stroke lesions on the DWI sequence for all patients, as patients without visible stroke lesions were excluded. Of the 51 patients, 16 were determined to have a mismatch in ischemic stroke signal intensities between the DWI sequence and the FLAIR sequence. All studies were annotated on an image level according to patient reports.

\section{Experiments and Results}

\subsection{Experimental Setup}

All experiments used one 40GB Nvidia A100 Tensor Core GPU for model training. The model architecture largely follows the implementation in \cite{Ronneberger2015}. The architecture has a depth of five and incorporates padded convolution, instance normalization, and ReLU activations. 

Models were trained for 500 epochs with a batch size of 2 and 4-step gradient accumulation on randomly extracted $128\times 128\times 128$ patches, with deep supervision, combining cross-entropy and generalized dice loss, and using an Adam optimizer with an initial learning rate of 0.0001 and a polynomial-rate schedule. 

We trained two U-Net models: a baseline U-Net trained without local gamma augmentation, and a U-Net trained with local gamma augmentations applied to DWI sequences, which we call U-Net+LG. Augmentations for \emph{both} models include: Rician noise, Gaussian noise, Gaussian blurring, contrast transformations, brightness transformations, global gamma correction, bias field correction, histogram equalization, elastic deformation, rotation, scaling, mirroring, random channel shifts and ghosting.

Of note is that a local gamma augmentation scheme as described in Section \ref{lga} does not account for images without a pathology.  Such images are paired with a segmentation map $M(x)$ which is identically zero. Such an $M(x)$ reduces Equation \ref{eqn-lga} to the identity map. To avoid this reduction for such images, we set $M(x) = 1$, so that a global gamma transformation can be applied to the patch.

\subsection{Evaluation}

We used image-level sensitivity and image-level specificity to evaluate the performance of a baseline U-Net and a local gamma augmented U-Net+LG on an ischemic stroke classification task. As segmentation models, U-Nets are typically evaluated with pointwise metrics such as the Dice-Sørensen coefficient or the Jaccard index. Our method of model evaluation, namely classification instead of segmentation, is required due to the lack of segmentation maps as annotations for the data set. To treat a segmentation model as a classification model, we define a U-Net prediction to be positive if at least one voxel in the model's output is predicted as corresponding to ischemic stroke.

\subsection{Results}

Table \ref{tab:image-metrics} summarises our results. We observe an increase in image-level sensitivity from 0.781 to 0.844 for the MVS dataset and an increase in image-level sensitivity from 0.757 to 0.814 for the RH dataset. These increases are accompanied with small reductions in image-level specificity on both datasets. 

For the WUS dataset, the improvement is even greater: an increase in image-level sensitivity from 0.566 to 0.736. We speculate that this larger increase in sensitivity is due to the influence of local gamma augmentation. In particular, the local gamma augmentation scheme exposes U-Net+LG to more examples of MRIs with signal differences between the DWI sequence and the remaining sequences. We also mention that specificity is not reported for WUS: a specificity score for this dataset is meaningless since every image contains an ischemic stroke.

\subsection{Discussion}
The simplicity of local gamma augmentation, namely its use of a one-parameter family of intensity transformations for data augmentation, means that it is very straightforward to implement and hence is amenable to further exploration. One potential research avenue would be to identify better choices of the gamma augmentation parameter. Our method uses a mixture distribution to select augmentation parameters, but this distribution is fixed. In this way, the chosen augmentation parameters are data-independent. Other research avenues could pursue the use of targeted gamma augmentations for other applications, such as adversarial training or pseudo-healthy synthesis.

A serious limitation of the method is its dependence on pointwise labels: without segmentation maps, only image-level gamma augmentations can be applied. Another limitation is the dependence on pathological tissues which are characterized by intensity differences relative to normal tissue. Not all brain diseases are characterized by intensity differences. Hence it is questionable that local gamma augmentation (or local intensity augmentation of any kind) would be effective.

\section{Conclusion}
We presented local gamma augmentation, an intensity-based method of data augmentation which, in a supervised setting, can be used to target pathological tissues in human brain MRIs. We applied this method to compensate for an intensity bias in the training set due to an over-representation of non-hyperacute ischemic strokes. This method was compared against a baseline data augmentation method without local gamma augmentation on three MRI data sets. The results suggest that restricting gamma transformations to ischemic stroke lesions during training of a segmentation model can enhance image-level sensitivity without impairing image-level specificity.

\printbibliography

\end{document}